\documentstyle[prd,tighten,aps]{revtex}

\begin{document}
\draft

%begin wide text
\twocolumn[\hsize\textwidth\columnwidth\hsize\csname
@twocolumnfalse\endcsname

\title{Asymptotically anti-de Sitter wormholes}

\author{Carlos Barcel\'o$^a$,
Luis J. Garay$^{b}$, Pedro F. Gonz\'alez-D\'{\i}az$^c$ and
Guillermo A. Mena Marug\'an$^c$}
\address{$^{(a)}$ Instituto de Astrof\'{\i}sica de
Andaluc\'{\i}a, CSIC, Camino Bajo de Hu\'etor, 18080 Granada,
Spain\\
$^{(b)}$ Theoretical Physics Group, The Blackett
Laboratory, Imperial College, London SW7 2BZ, UK\\
$^{(c)}$ Centro de F\'{\i}sica ``Miguel Catal\'an'',
Instituto de Matem\'aticas y F\'{\i}sica Fundamental,
CSIC,\\
 Serrano 121, 28006 Madrid, Spain}
\date{23 October 1995}
\maketitle

%preprint number
\vspace*{-4.5cm} \begin{flushright}
gr-qc/9510047, Imperial/TP/95--96/06
\end{flushright} \vspace*{3.3cm}

\begin{abstract}

Starting with a procedure for dealing  with general asymptotic
behaviors, we construct a quantum theory for asymptotically
anti-de Sitter wormholes. We follow both the path integral
formalism and the algebraic quantization program proposed by
Ashtekar. By adding suitable surface terms, the Euclidean action
of the asymptoically anti-de Sitter wormholes can be seen to be
finite and gauge invariant. This action determines an
appropriate variational problem for wormholes. We also obtain
the wormhole wave functions of the gravitational model and show
that all the physical states of the quantum theory are
superpositions of wormhole states.

\end{abstract}


%end wide text
\vskip2pc]

\renewcommand{\theequation}{\arabic{section}.\arabic{equation}}

\section{Introduction}

Wormholes are topology changes that connect different regions of
spacetime which may be far apart \cite{ha90,LI}.
In the dilute wormhole approximation \cite{ha90,CC}, these
regions are regarded as asymptotically large.  Wormholes can be
represented by quantum states, i.e. solutions of the
Wheeler-DeWitt equation (and the quantum momentum constraints),
which satisfy some suitable boundary conditions on the
asymptotic regions \cite{HP,L1}.
They can also be considered as instantons, solutions of the
Euclidean Einstein equations, which join the two asymptotic
regions of spacetime by a throat \cite{LI,CWH,HL}.  As saddle
points of the Euclidean action, these instantons would allow the
Euclidean path integral to be approximated semiclassically, thus
representing quantum tunnelling effects between the asymptotic
regions.

Asymptotically flat wormholes have been extensively studied
in the literature \cite{flat}.  There exist however other asymptotic
behaviors \cite{HL,L2,MY,LID} that are worth considering.
For instance, wormholes
whose asymptotic regions are Kantowski-Sachs spacetimes
\cite{L2}, with the topology of $I\!\!R^3\times S^1$, may
provide a link between black hole physics and the issue of
topology change. Asymptotically anti-de Sitter wormholes are also
of particular interest.  In this case, the asymptotic regions
expand exponentially (in proper time) due to the presence of an
effective negative cosmological constant.  These wormholes could be
regarded as excited states  in the sense that the
cosmological constant could be interpreted as a non-vanishing
asymptotic energy  of the matter fields.  On the other
hand, one should expect that these wormholes could give  a
non-vanishing contribution to the path integral and,
consequently, they should be taken into account in calculations
such as those leading to Coleman's mechanism for the vanishing
of the effective cosmological constant
\cite{CC}.

It has been argued that wormholes might affect the constants of
nature through low energy effective interactions
\cite{LI,CC,cole}.  The existence of a Hilbert structure in the
space of wormhole wave functions is essential to turn the
apparent non-local interaction introduced by wormholes into a
local one, as seen from one of the asymptotic regions
\cite{LI,CC,cole,giddst}.  Such Hilbert space structure is therefore
necessary in the explicit calculation of these effective
interactions.

In this work, we construct the Hilbert space of asymptotically
anti-de Sitter wormholes, suggesting a procedure for dealing with
other possible asymptotic behaviors.  We employ the path
integral approach to obtain the quantum states and Ashtekar's
algebraic program \cite{ASH} to complete the quantization of these
wormholes, including the determination of the physical inner
product.  Finding a well-defined set of wormhole boundary
conditions becomes a central  issue in both approaches.

Hawking and Page \cite{HP} have proposed that the boundary
conditions for the quantum wormhole states should guarantee that
their corresponding wave functions are exponentially damped for
large three-geometries, so that one recovers the semiclassical
behavior expected in the asymptotic limit of large Euclidean
configurations. Besides, the wormhole wave functions should be
regular for all regular matter fields and three-geometries,
including those geometries that degenerate to zero because of an
ill-defined slicing of  spacetime.  From the path integral point
of view, these conditions can be accomplished if the wormhole
wave functions are defined by the sum over all possible
spacetimes with the prescribed asymptotic behavior and over all
matter fields that are compatible with the given asymptotic
spacetime via the vanishing of the first class constraints  in
the asymptotic regions.  For instance, if we are dealing with
asymptotically flat spacetimes, the energy-momentum tensor of
the matter fields will have to vanish at infinite proper time
\cite{HP,L1},
or if an asymptotically anti-de Sitter behavior is considered,
then the matter content will have to induce an effective
negative cosmological constant in the asymptotic region.  As a
previous step,  we implement the wormhole boundary conditions
canonically and find an appropriate gauge invariant action,
which is finite for classical wormhole solutions. This amounts
to include the surface terms that are characteristic of
asymptotic spacetimes (see Ref. \cite{L1,gibbs}) and
that remove the infinite contribution of the asymptotic regions.

In order to determine the Hilbert structure of the space of
wormholes, and thus reach a consistent quantum theory to
describe these states, we follow the algebraic quantization
program put forward by Ashtekar \cite{ASH}. In the following,
we briefly summarize the main steps of this program.  One must
first choose a complete set of classical variables that is
closed under Poisson brackets and complex conjugation.  To each
of these elementary variables one associates an abstract
operator and constructs the algebra generated by them, imposing
on it the canonical commutation relations. One must next find a
linear representation of this algebra on a complex vector space
and choose explicit operators to represent the first-class
constraints of the system.  The subspace annihilated by these
constraints supplies the space of quantum states, and the
quantum observables are the operators that leave this space
invariant \cite{ASH}.  The physical inner product on quantum
states can then be determined by requiring that the complex
conjugation relations between elementary variables (usually
called reality conditions) are realized as Hermitian adjoint
relations between quantum observables on the resulting Hilbert
space \cite{INV}.  Actually, if an inner product satisfying this
condition exists, it is unique under very general assumptions
\cite{REN}.  The elements in the Hilbert space obtained in this
way are the physical states of the theory.

For gravitational systems which exhibit quantum wormhole
solutions, if one chooses properly the representation space,  it
is possible to show that the space of quantum states coincides
with that spanned by the wormhole wave functions, provided that
the latter is invariant under the action of the quantum
observables \cite{BAS}. Therefore, the inner product of
wormholes can in fact be determined by imposing an adequate set
of reality conditions, and the corresponding Hilbert space of
wormholes can be identified with that of physical states of the
quantum theory.

In section II, we present a model which illustrates the general
features discussed above.  It consists of a scalar field
conformally coupled to a homogeneous and isotropic spacetime
with a negative cosmological constant.  In section III, we show
that such a model possesses asymptotically anti-de Sitter wormhole
solutions.  In section IV, an appropriate action  for
asymptotically large spacetimes is constructed in the general
context of superspace  and particularized then to our
minisuperspace model. The path integral quantization is
discussed in section V.  Using the results of this section, we
carry out the full algebraic quantization of the model in
section VI.  We finally summarize and conclude in section VII.

\section{The model}
\setcounter{equation}{0}

We shall discuss in detail a homogeneous and isotropic
gravitational minisuperspace model provided with a conformally
coupled scalar field and a negative cosmological constant.  As
we shall see in Sec. III, this model possesses instanton
solutions which are asymptotically anti-de Sitter.

We  start by performing the standard 3+1 splitting of the
Euclidean spacetime metric
\begin{equation}
ds^{2}=(N^{2}+N^{i}N_{i})d\tau^{2}+2N_{i}d\tau
dx^{i}+g_{ij}dx^{i}dx^{j}
\end{equation}
where $N$ and $N^i$ are the lapse and shift functions and
$g_{ij}$ is the metric on the closed three-surfaces of constant
time. The Euclidean action can be written in the Hamiltonian
form
\begin{equation}
\tilde{I}=\int d\tau\int d^3x
[\pi^{ij}\dot{g}_{ij}+\pi_{\phi}\dot{\phi}-N{\cal H}-N^{i}{\cal
H}_{i}],
\label{tildeI}
\end{equation}
in which $\pi^{ij}$ and $\pi_\phi$ are the canonical momenta
conjugate to the three-metric $g_{ij}$ and the conformal scalar
field $\phi$, and the dot denotes the derivative with respect to
the time coordinate $\tau$. In the above expression, ${\cal H}$
and ${\cal H}_{i}$ are the standard ADM Hamiltonian and momentum
constraints for Euclidean gravity conformally coupled to a
scalar field in the presence of a negative cosmological constant
$\Lambda$.

The requirements of homogeneity and isotropy, i.e. the
restriction to the minisuperspace under consideration, can be
imposed by writing the spacetime metric in the form
\begin{equation}
ds^{2}=\frac{2G}{3\pi}[N^2(\tau)d\tau^{2}+a^2(\tau)
\Omega_{ij}dx^idx^j],
\label{metric}
\end{equation}
being $\Omega_{ij}$ the metric on the unit three-sphere and $G$
Newton's constant; likewise, the scalar field will depend only
on the time coordinate, $\phi=\phi(\tau)$. It is
convenient to introduce a new variable $\chi$ to describe the
conformal scalar field in the following manner
\begin{equation}
\phi=\sqrt{\frac{3}{4\pi G}}\; \frac{\chi}{a}.
\end{equation}

When particularized to this minisuperspace model, the Euclidean
action becomes
\begin{equation}
\tilde{I}=\int d\tau [\pi_{a}\dot{a}+
\pi_{\chi}\dot{\chi}-NH].
\label{tildeImss}
\end{equation}
Here, ($\pi_{a}$,$\pi_{\chi}$) are the momenta canonically
conjugate to the variables ($a$,$\chi$), and are related to the
superspace canonical momenta ($\pi^{ij},\pi_{\phi}$) through the
formulas
\begin{eqnarray}
\pi^{ij}&=&\frac{1}{8\pi G}
\left(\frac{\pi_{a}}{a}+\frac{\pi_{\chi}\chi}{a^{2}}\right)
\Omega^{ij}\Omega^{1/2},\\
\pi_{\phi}&=&\sqrt{\frac{G}{3\pi^3}}\;\
a\pi_{\chi}\Omega^{1/2},
\end{eqnarray}
with $\Omega=\det \Omega_{ij}$.  On the other hand, $H$ denotes
the Hamiltonian constraint in minisuperspace, namely
\begin{equation}
H=\frac{1}{2a}(-\pi_{a}^{2}+a^2+\lambda a^{4}+
\pi_{\chi}^{2}-\chi^{2}),
\end{equation}
where $\lambda=-\frac{2G}{9\pi}\Lambda >0$.

\section{Classical Solutions}
\setcounter{equation}{0}

The classical Euclidean solutions of this model can be easily
obtained by introducing the conformal time $d\eta=d\tau/a$. If
we denote the derivative with respect to this time by a prime,
the dynamical equations read
\begin{eqnarray}
&a^{\prime}=-\pi_{a},
\hspace{10mm}
&\pi_{a}^{\prime}=-a-2\lambda a^{3} ,
\label{class1}
\\
&\chi^{\prime}=\pi_{\chi},
\hspace{13mm}
&\pi_{\chi}^{\prime}=\chi,
\label{class2}
\end{eqnarray}
while the Hamiltonian constraint is
\begin{equation}
\frac{1}{2}\left(-\pi_{a}^{2}+a^{2}+\lambda a^{4}+\pi_{\chi}^{2}-\chi^{2}
\right)=0.
\label{constr}
\end{equation}
In the above expressions, we have set the lapse function equal
to one.

The general solution to Eqs. (\ref{class2}) is given by
\begin{equation}
\chi=A\cosh\eta+B\sinh\eta,
\end{equation}
with $A$ and $B$ being two arbitrary real constants.
Substituting this solution in the Hamiltonian constraint and
using the first equation in (\ref{class1}), we get
\begin{equation}
(a')^{2}=a^{2}+\lambda a^{4}-2E,
\label{whconstr}
\end{equation}
where $E=\frac{1}{2}(A^2-B^2)$.  This constraint will have
solutions of the wormhole type only if the polynomial that
appears on its right hand side has at least a positive root.
This implies that $E$ must be positive.  We will restrict to
this case hereafter.

Since $E>0$, we can parametrize the constants $A$ and $B$ as
\begin{equation}
A=\sqrt{2E}\cosh\eta_{0},
\hspace{8mm}
B=-\sqrt{2E}\sinh\eta_{0},
\end{equation}
with $\eta_{0}$ an arbitrary real parameter. The conformal field
$\chi$ can then be rewritten
\begin{equation}
\chi=\sqrt{2E}\cosh(\eta-\eta_{0}).\end{equation}
Besides, integration of Eq. (\ref{whconstr}) leads to
\begin{equation}
a(\eta)=a_M\;\mbox{nc}(D^{1/4} (\eta-\tilde{\eta}_{0})\;|m),
\label{aconf}
\end{equation}
where $\mbox{nc}(u|m)$ is the Jacobian elliptic function with
parameter $m$ \cite{JAC}, $\tilde\eta_{0}$ is a real constant,
and
\begin{equation}
D=1+8\lambda E,
\hspace{1cm}
a_M=\left(\frac{D^{1/2}-1}{2\lambda}\right)^{1/2},
\label{classpar}
\end{equation}
%\vspace{-5mm}
\begin{equation}
m=\frac{D^{-1/2}+1}{2}.
\label{classpar2}
\end{equation}
One can check that Eqs. (\ref{class1}) are then
straightforwardly satisfied.

The classical wormhole solutions of the model are therefore
parametrized by three independent real constants: $\eta_{0}$,
$\tilde\eta_{0}$ and $E>0$.  Notice that $D>1$ and that $a_M$ is
the size of the wormhole throat, which coincides with the only
positive root of the right hand side of the constraint
(\ref{whconstr}).

It is also possible to obtain the solution to that constraint in
terms of the proper time $\tau$. One arrives at the following
expression for the scale factor
\begin{equation}
a=\frac{1}{\sqrt{2\lambda}}\left\{ D^{1/2}\cosh\left[2\sqrt\lambda(\tau-
\tilde{\tau}_{0})\right]-1\right\}^{1/2}    ,
\label{eq:factor}
\end{equation}
where the new real constant $\tilde{\tau}_0$ appears instead of
$\tilde \eta_0$.

Some comments are in order at this point. Firstly, the conformal
time $\eta$ tends to a finite value $\eta_{M}$ as the proper
time $\tau$ goes to infinity.  This is due to the fact that,
being the scale factor exponentially large at
$\tau\rightarrow\infty$, the integral
$\int^{\infty}d\tau/a(\tau)$ converges. This feature is actually
reflected by the elliptic function $\mbox{nc}(u|m)$ that
describes the scale factor solutions in conformal time, for such
a function diverges at the finite point $u=K(m)$, with $K(m)$ being
the complete elliptic integral of the first kind \cite{JAC}.
Secondly, all the solutions that we have obtained are
asymptotically anti-de Sitter, as can be easily seen by
considering the limit $\tau\rightarrow\infty$ in Eq.
(\ref{eq:factor}).  The globally anti-de Sitter solution
corresponds to the limit $D\rightarrow1$ in that equation. Finally, note
that the flat solutions ($\lambda=0$) cannot be recovered by
taking the limit $\lambda\rightarrow 0$. This is not surprising,
because the $\lambda$-term in Eq. (\ref{whconstr}) is dominant
in the asymptotic region $a\rightarrow\infty$ and therefore
provides a singular perturbation to the $\lambda=0$ equations of
motion.

\section{Surface Terms}
\setcounter{equation}{0}

Action (\ref{tildeImss}) is not adequate for studying spacetimes
that join onto an asymptotically anti-de Sitter region. Actually,
it diverges for classical solutions \cite{MY} and can be shown
not to be invariant under time reparametrizations that map the
initial three-surface onto itself. Moreover, it is not quite
clear that this action could correspond then to a variational
problem which guaranteed the anti-de Sitter asymptotic behavior
of the classical spacetimes. These difficulties can be none the
less overcome by adding appropriate surface terms to the action.
In order to obtain these terms, it appears most convenient to begin
by considering the general superspace framework, without
specialising to any particular asymptotic behavior. We shall
then reduce to the homogeneous and isotropic model conformally
coupled to a scalar field, discussing first the flat case
$\lambda=0$ to circumvent the subtleties that arise
when introducing a negative cosmological constant.

\subsection{Superspace}

The gravitational systems under consideration join an initial
three-surface onto an asymptotic region. The boundary conditions
for the associated variational problem
must reflect this fact. The geometry of the initial
three-surface and its matter content will be chosen as one of
the boundary conditions. The final time boundary conditions must
guarantee the prescribed asymptotic behavior (at least for
classical solutions). Besides, we would like our system to be invariant
under gauge transformations that are not fixed at the final
time, so that one can reach a semiclassical picture in which the
final surface is not fixed, but asymptotically embedded in a
classical spacetime.

Let us assume that the final boundary conditions can be imposed
by fixing certain variables $Q^\alpha$ at the final time
$\tau_f$, namely $Q^\alpha\big|_{\tau_f}=Q_f^\alpha$.  Notice
that the proper time goes to infinity when
$\tau\rightarrow\tau_f$ for the models studied so far
\cite{LI,L1,CWH,HL}. In terms of these new variables $Q^\alpha$ and
their canonically conjugate momenta $P_\alpha$ the action
(\ref{tildeI}) acquires the form
\begin{eqnarray}
\tilde I&=&\int_0^{\tau_f} d\tau\int d^3x
\left(P_\alpha\dot Q^\alpha
-N{\cal H} -N^i{\cal H}_i  \right)
\nonumber\\
&&+\int d^3x\; {\cal F}\big|_{\tau_f}-\int d^3x\; {\cal
F}\big|_0,
\end{eqnarray}
where ${\cal F}={\cal F}[g_{ij},\phi|Q^\alpha]$ is a generating
functional for the canonical transformation from the
geometrodynamical variables to $(Q^{\alpha},P_{\alpha})$.  Then,
it can be seen that the action
\begin{equation}
I=\tilde I -\int d^3x\;{\cal F}\big|_{\tau_f}
\label{actionI}
\end{equation}
is appropriate for fixing the initial three-geometry, the
initial scalar field and the asymptotic variables $Q^\alpha$.

As mentioned above, this action should be invariant under
spatial diffeomorphisms and time reparametrizations that are
restricted only to map the initial surface $(\tau=0)$ onto
itself. These transformations are generated by $\cal H$ and
${\cal H}_i$ via the standard Poisson bracket relations $
\delta A=\left\{A, \int d^3x\left( \epsilon {\cal H}
-\epsilon ^i{\cal H}_i  \right)\right\}, $ with $\epsilon$
vanishing at $\tau=0$.  The variation of the action $I$ under
these transformations is
\begin{equation}
\delta I=-\int d^3x\left(\epsilon
{\cal H}+\epsilon^i {\cal H}_i- P_\alpha \delta Q^\alpha
\right)\big|_ {\tau_f},\label{vari}
\end{equation}
where we have used the standard gauge variation for the lapse
and shift functions \cite{teitel}.  Since the gauge transformations
are arbitrary at the final time, the vanishing of the first two
terms in the right hand side of this expression is only ensured
by choosing the variables $Q^\alpha$ so that the first class
constraints are set to zero in the asymptotic region:
\begin{equation}
{\cal H}\big|_{Q^\alpha_f}=0,
\hspace{15mm}
{\cal H}_i\big|_{Q^\alpha_f}=0.
\end{equation}
The values $Q^\alpha_f$ cannot therefore be fixed in a fully
arbitrary way.  For the vanishing of the third term in
(\ref{vari}), on the other hand, we need our canonical
coordinates $Q^\alpha$ to be locally observable in the
asymptotic region, in the sense that the Poisson brackets
$\left\{Q^\alpha,\cal H
\right\}\big|_{Q^\alpha_f}$ and $\left\{Q^\alpha,{\cal H}_i
\right\}\big|_{Q^\alpha_f}$ vanish, so that their asymptotic values are
left invariant under the gauge transformations of the system.

The resulting action $I$ turns out to be finite for classical
solutions under sufficiently general conditions.  To see this we
first note that, on classical solutions,
\begin{eqnarray}
I_{\rm class}&=&\int_0^{\tau_1} d\tau \int d^3x
\left(\pi^{ij}\dot g_{ij}+
\pi_\phi \dot \phi \right)- \int d^3x {\cal F}\big|_{\tau_1}
\nonumber \\
&&+\int_{\tau_1}^{\tau_f} d\tau \int d^3x
\left(\pi^{ij}\dot g_{ij}+
\pi_\phi \dot \phi -\dot {\cal F}\right),
\label{Iclass}
\end{eqnarray}
where $\tau_1$ is a finite intermediate time.  Since the
classical solutions should be regular along the entire interval
$[0,\tau_1]$ but might blow up asymptotically as $\tau$
approaches $\tau_f$, any possible divergence in (\ref{Iclass})
must appear in the last integral.  Taking into account the
canonical transformation generated by ${\cal
F}[g_{ij},\phi|Q^\alpha]$, we rewrite this last integral as
\begin{equation}
\int_{\tau_1}^{\tau_f} d\tau \int d^3x
P_\alpha \dot Q^\alpha .
\label{finiteclass}
\end{equation}
If the variables $Q^\alpha$ are actually observables, i.e. if their
Poisson brackets with the constraints vanish weakly, integral
(\ref{finiteclass}) vanishes, because these variables are then
constant on the classical trajectories. In the more general case
in which they are only locally observable at their asymptotic
values, $\dot Q^\alpha\rightarrow 0$ as we approach $\tau_f$,
and  the action will be finite if the term $\int d^3x P_\alpha
\dot Q^\alpha $ decreases fast enough in the limit
$\tau\rightarrow \tau_f$.  This further restricts the kind of
variables that are allowed to be fixed asymptotically.

To summarize, the asymptotic boundary conditions can be
canonically implemented  by choosing a suitable set of
compatible variables and fixing their final values  in such a
way that they become locally observable.  These values must
imply, in particular, the asymptotic vanishing of the generators
of spatial diffeomorphisms and time reparametrizations.  This
procedure ensures that the action for the system is gauge
invariant, finite and gives rise to a well-defined variational
problem for the boundary conditions under consideration.

\subsection{Asymptotically flat wormholes}

We first consider the case of asymptotically flat spacetimes
($\lambda=0$) \cite{L1} for which action (\ref{tildeImss})
can be rewritten as
\begin{equation}
\tilde{I}=\int_{0}^{\eta_f}d\eta\;[\pi_{a}a^{\prime}+
\pi_{\chi}\chi^{\prime}-NH],
\end{equation}
where $\eta$ is again the conformal time, $\eta_f=\infty$, and
the Hamiltonian constraint $H$ is the difference of the
Hamiltonians of two harmonic oscillators, one  describing the
scale factor and the other the conformal field.

We expect  the wormholes solutions of this model to be
stationary trajectories of the variational problem with fixed
initial values of $a$ and $\chi$ and suitable final values for a
complete set of compatible variables which are left invariant
under time reparametrizations.  These conditions on the
variables fixed in the asymptotic region will be clearly
satisfied if they are compatible observables of the system.

Given the form of the Hamiltonian constraint, we can choose
\begin{equation}
E_a=\frac{1}{2}(a^{2}-\pi_{a}^{2}) ,
\hspace{10mm}
E_{\chi}=\frac{1}{2}(\chi^{2}-\pi_{\chi}^{2})
\end{equation}
as our set of compatible observables. The variables
\begin{equation}
\Theta_x=\ln{\left(\frac{x+\pi_x}{\sqrt{x^2-\pi_x^2}}
\right)}\hspace{10mm}(x=a,\chi)
\end{equation}
are the momenta canonically conjugate to these observables.
The canonical transformation from $(x,\pi_x)$ to
$(E_x,\Theta_x)$ is generated by the function
\begin{eqnarray}
&&F_{x}(x|E_x)=-\int^{x}_{\sqrt{2E_x}}dz(z^{2}-2E_{x})^{1/2}
\nonumber\\
&&=-\frac{x}{2} (x^{2}-2E_{x})^{1/2}
%\nonumber\\
%&
+ E_{x} \ln{\left(\frac{x+\sqrt{x^2-2E_x}}{\sqrt{2E_x}}
\right)}.\label{A.5}
\end{eqnarray}
In terms of the new variables, action (\ref{actionI}) reduces to
\begin{eqnarray}
I&=&\tilde{I}-(F_{a}+F_{\chi})\big|_{\eta_f}
\nonumber\\
&=&\int_{0}^{\eta_f}d\eta\;[\Theta_{a}E^{\prime}_{a}+\Theta_{\chi}
{E}^{\prime}_{\chi}-N(E_{a}-E_{\chi})]\nonumber\\
&&-(F_{a}+F_{\chi})\big|_{0}\ ,\label{A.6}
\end{eqnarray}
with $\eta_f=\infty$.  On the other hand, the Hamiltonian
constraint $ H =E_a-E_{\chi}$ generates, via Poisson brackets,
the time reparametrizations:
\begin{equation}
\delta E_x=\epsilon\{E_x,H\},\hspace{.4cm}
\delta \Theta_x=\epsilon\{\Theta_x,H\},
\hspace{.4cm}
\delta N=\epsilon^{\prime},
\end{equation}
where the parameter $\epsilon$ depends only on the conformal
time.  It is then  easy to check that the
action (\ref{A.6}), supplemented with the wormhole boundary
conditions
\begin{equation}
E_a(\eta_f)=E_{\chi}(\eta_f)=E, \hspace{5mm}
\mbox{with $E>0$},
\label{A.10}
\end{equation}
is invariant under time reparametrizations that map the initial
surface onto itself (namely with $\epsilon(0)=0$).  The
stationary points of this action are the classical trajectories
that join an initial three-surface characterized by the scale
factor $a(0)=a_i$ and the conformal field $\chi(0)=\chi_i$ with
an asymptotic region in which condition (\ref{A.10}) is
satisfied. This asymptotic condition actually implies that the
solutions of the model are asymptotically flat, as can be
straightforwardly seen by solving the equation
$2E=a^2-(a^{\prime})^2$.  Finally,
given the constraint $H=0$ and the dynamical equations
$E_a^{\prime}=E_{\chi}^{\prime}=0$, the action (\ref{A.6})
reduces  to
\begin{equation}
I_{\rm class}=-F_a(a_i|E)-F_{\chi}(\chi_i|E)
\label{A.11}
\end{equation}
on classical solutions. From Eq. (\ref{A.5}), it then follows
that the classical action is always finite provided that $E$
(i.e. the asymptotic energy of the conformal field) is positive.

\subsection{Asymptotically anti-de Sitter wormholes}

Let us now extend the above analysis to the asymptoticaly
anti-de Sitter case.  The situation remains in fact unchanged
except in what refers to the scale factor. In the anti-de Sitter
case, the part of the Hamiltonian costraint which depends on $a$
and $\pi_a$ incorporates a cosmological term, namely
$E_a=\frac{1}{2}(a^2+\lambda a^4-\pi_a^2)$.  The generating
function $F_a(a|E_a)$ has to be subsequently modified to take
care of the non-vanishing cosmological constant. One arrives at
\begin{equation}
F_{a}(a|E_a)=-\int^{a}_{a_M}dz(z^{2}+\lambda
z^{4}-2E_{a})^{1/2},
\label{A.12}\end{equation}
where $a_M$ is the root of the polynomial $a^2+\lambda
a^4-2E_{a}$ which can be obtained from Eq. (\ref{classpar}) by
substituting $E_a$ for $E$.

Expressions (\ref{A.6}) and (\ref{A.10}) still provide the
gauge invariant action
 and the  boundary conditions for
the anti-de Sitter wormholes, respectively. Note however that,
from our remarks at the end of Sec. III, the final conformal
time $\eta_f$ will now be finite for all the wormhole solutions
of the model.  We shall therefore fix $\eta_f$ to coincide with
the time $\eta_M(a_i,E)$ at which the solution
(3.8--10), verifying $a(0)=a_i$, tends to
$+\infty$. Finally, one can check that the action on classical
solutions again takes the form (\ref{A.11}), but with
$F_a(a|E_a)$ supplied now by Eq. (\ref{A.12}).

\section{Path Integral}
\setcounter{equation}{0}

The path integral which provides the anti-de Sitter quantum
wormholes parametrized by the asymptotic value of the conformal
field energy $E>0$ is given by
\begin{eqnarray}
\Psi_{E}[a_{i},\chi_{i}]
&=&\int {\cal D}N{\cal D}
\mu(a,\pi_{a},\chi,\pi_{\chi})\Delta_{FP}\delta(N-1)\nonumber\\
&&\times\exp[-I(a,\pi_{a},\chi,\pi_{\chi},N)].
\end{eqnarray}
Here, we sum over histories satisfying $a(0)=a_{i}$,
$\chi(0)=\chi_{i}$ and $E_a(\eta_{M})=E_\chi (\eta_{M})=E$.  We
recall that $\eta_M$ is a constant that depends on the values of
$a_i$ and $E$.  The Faddeev-Popov determinant $\Delta_{FP}$ can
be set equal to the unity, because it does not depend on any of
the integration fields for our gauge fixing condition $N=1$.
Integration over $N$ leads then to
\begin{equation}
 \Psi_{E}[a_{i},\chi_{i}]=\int {\cal D}\mu(a,\pi_{a},\chi,
\pi_{\chi})\;\exp(-I),
\label{pint}
\end{equation}
where
\begin{eqnarray}
I=\int_{0}^{\eta_{M}} d\eta\;
\big[&&\pi_{a}a^{\prime}+\pi_{\chi}\chi^{\prime}
-\frac{N}{2}(-\pi_{a}^{2}+a^{2}+\lambda a^{4} \nonumber \\
&&+\pi_{\chi}^{2}-\chi^{2})\big]-(F_{a}+F_{\chi})\big|_
{\eta_{M}},
\end{eqnarray}
The part of this path integral which depends on the conformal
field provides the propagator $U(E,\eta_M|\chi_i,0)$ of a harmonic
oscillator between a fixed initial field $\chi_i$ and a constant
energy $E_{\chi}=E$ at the final time $\eta_M$. With a proper
choice of the integration measure, this propagator would be a
linear combination of the normalized eigenstates
$\varphi_n(\chi_i)\ (n=0,1\ldots)$ of the harmonic oscillator,
namely
\begin{equation}
U(E,\eta_M|\chi_i,0)=\sum_{n=0}^{\infty}
e^{-\eta_{M}(n+\frac{1}{2})}\omega_{n}(E)
\varphi_{n}(\chi_{i}),
\label{hosc}
\end{equation}
in which $\omega_{n}(E)$ are some coefficients which depend on
$E$ and we have set $\hbar=1$.  On the other hand, the result of
the path integral should satisfy the quantum version of the
constraint
\begin{equation} -\pi_{\chi}^2+\chi^2-2E=0\end{equation}
which, since $E_{\chi}$ is preserved by the dynamics of the
system and we have imposed $E_{\chi}=E$ at $\eta_M$, holds on
all classical trajectories.  Therefore using Eq. (\ref{hosc}), we
conclude that $E$ can only take the values
$n+\frac{1}{2}$, if the path integral is to be well-defined, and
then that,  up to a global $E$-dependent factor,
\begin{equation}
U(n+\frac{1}{2},\eta_M|\chi_i,0)=e^{-\eta_M(n+\frac{1}{2})}
\varphi_n(\chi_i).\end{equation}
Hence,  the path integral  reduces  to
\begin{equation}
\Psi_{n+\frac{1}{2}}[a_{i},\chi_{i}]=
\varphi_{n}(\chi_{i})\Phi_{n}(a_{i}),
\end{equation}
where
\begin{eqnarray}
&&\Phi_{n}(a_{i})=\int {\cal D}\mu(a,\pi_{a}) \;\exp\bigg\{-
\int_{0}^{\eta_{M}} d\eta\;\big[
\pi_{a}a^{\prime}\nonumber \\
&&-\frac{1}{2}(-\pi_{a}^{2}+a^{2}+
\lambda a^{4}-(2n+1))\big]+F_{a}\big|_{\eta_{M}}\bigg\}.
\label{pint2}
\end{eqnarray}
In this expression, we sum  over histories with $a(0)=a_{i}$ and
$E_{a}(\eta_{M})=n+\frac{1}{2}$.  The functions $\Phi_n(a)$
must be solutions to the Wheeler-DeWitt equation which follows
from the constraint
\begin{equation}
-\pi_{a}^{2}+a^{2}+\lambda a^{4}-(2n+1)=0. \label{eq:lign}
\end{equation}
The factor ordering in this Wheeler-DeWitt equation will depend
on the integration measure employed in the path integral
(\ref{pint2}).  We shall assume a factor ordering of the form
\begin{eqnarray}
\hat{H}_a\Phi_n(a)&\equiv&
\frac{1}{2}\left(-\frac{1}{f(a)}\partial_a
f(a)\partial_a+a^2+\lambda a^4\right)\Phi_n(a)\nonumber\\
&=&\left(n+\frac{1}{2}\right)
\Phi_n(a),
\label{wdwa}\end{eqnarray}
where the function $f(a)$ will be supposed to be analytic and
strictly positive at least for $a\geq 0$ and such that
\begin{equation}
\lim_{a\rightarrow\infty}\frac{f^{\prime}(a)}{a^2f(a)}=0,
\label{condlim}
\end{equation}
the prime denoting here the first derivative.

If we now restrict our attention to the region $a\in I\!\!R^+$, so
that each different geometry of the type (\ref{metric}) is
considered only once, it is possible to prove that there
actually exists a solution $\Phi_n(a)$ to Eq. (\ref{wdwa}) such
that it is regular in the positive semiaxis  and decreases
exponentially for large scale factor. In order to see this, let
us consider $\hat{H}_a-(n+\frac{1}{2})$ as a second order
differential operator which annihilates $\Phi_n(a)$. The
coefficient of $\partial_a^2$ in this operator is constant. The
coefficient of $\partial_a$, given by $f^{\prime}(a)/f(a)$, is
analytic in $a\geq 0$, because $f(a)$ is positive and analytic
in this semiaxis. Finally, the non-derivative term is also
analytic, as it is a polynomial in $a$.  It then follows
\cite{BP} that, for each fixed $n$, the differential equation
(\ref{wdwa}) possesses two linearly independent solutions which
are analytic at least for all $a\geq 0$.  Moreover, provided
that condition (\ref{condlim}) is satisfied, an asymptotic
analysis of this differential equation shows that one of these
solutions must be exponentially damped in the limit
$a\rightarrow \infty$, while the other increases exponentially.

We want to show now that $\Phi_{n}(a)$ should be the
exponentially damped solution.  For $a_{i}\gg 1$, we expect the
semiclassical aproximation to become valid in the path integral,
i.e.  $\Phi_{n}(a_{i})\sim e^{-I_{\rm class}}$ being $I_{\rm
class}$ the action of the classical solution to the constraint
(\ref{eq:lign}) with $a(0)=a_{i}$.  For this solution,
$a(\eta\rightarrow\eta_{M})\rightarrow\infty$ and, admitting
that $a^{\prime}=-\pi_a$ is positive for $a\gg 1$, one gets
\begin{eqnarray}
I_{\rm class}&=&-\int_{a_i}^{\infty}dz\;\left[z^{2}+\lambda
z^{4}-(2n+1)\right]^{1/2}-F_a\big|_{a=\infty} \nonumber \\
&=&\int_{a_M}^{a_i}dz\;\left[z^{2}+\lambda z^{4}-(2n+1)\right]^{1/2},
\end{eqnarray}
where we have substituted Eq. (\ref{A.12}), and $a_M$ is given
by Eq. (\ref{classpar}) with $E=n+\frac{1}{2}$.  The integral in
the above expresion is positive and diverges in the limit
$a_i\rightarrow\infty$. As a consequence, the function
$\Phi_n(a_i)$ is exponentially damped in that limit.

We thus conclude that the functions $\Phi_n(a)$, solutions to
(\ref{wdwa}) with $n=0,1\ldots$, satisfy the wormhole boundary
conditions if $a$ is restricted to run over the positive axis.
Actually, we have shown that these functions are not only
regular, but analytic in $ a\geq 0$.

It is worth remarking that, even though the solutions
$\Phi_n(a)$ could be analytically extended to the whole real
axis, their asymptotic behavior at $a\rightarrow -\infty$ would
not be damped unless in exceptional situations, and
never for all the functions $\Phi_n(a)$ ($n=0,1\ldots$), because
that would imply that the operator $\hat{H}_a$  has exactly the
eigenvalue spectrum which characterizes the Hamiltonian of the
harmonic oscillator. Therefore, the restriction to $a\in
I\!\!R^+$ is essential if we want that the wave functions
$\Phi_n(a)$ represent quantum wormhole states.

\section{Algebraic Quantization}
\setcounter{equation}{0}

Our minisuperspace model possesses only one constraint, namely
the Hamiltonian contraint (\ref{constr}). To carry out the
algebraic quantization, it is convenient to introduce the
Lorentzian momenta $(\Pi_a,\Pi_{\chi})$ canonically conjugate to
the scale factor and the conformal field. Then, the Hamiltonian
constraint reads
\begin{equation}
H=\frac{1}{2}(\Pi_a^2+a^2+\lambda
a^4)-\frac{1}{2}(\Pi_{\chi}^2+\chi^2)=0.
\label{Hlor}
\end{equation}

The symplectic structure on phase space is supplied by the
Poisson brackets $\{a,\Pi_a\}=1$ and $\{\chi,\Pi_{\chi}\}=1$.
For Lorentzian geometries and real conformal fields, we have
$\chi,\Pi_a,\Pi_{\chi}\in I\!\!R$. Besides, we shall
restrict the scale factor to be positive, $a\in I\!\!R^+$,
so that each different four-geometry is considered only once.

\subsection{Elementary variables}

As pointed out in the introduction,
our first task will consist in choosing a suitable complete set
of elementary variables in the phase space of the model.  Since
the part of the Hamiltonian constraint  which depends on the
conformal field can be interpreted as the Hamiltonian of a
harmonic oscillator, we will describe the degrees of freedom of
this field by the annihilation and creation variables
\begin{equation}
A_{\chi}=\frac{1}{\sqrt{2}}(\chi+i\Pi_{\chi}),
\hspace{7mm}
A^{\dag}_{\chi}=\frac{1}{\sqrt{2}}(\chi-i\Pi_{\chi}).
\end{equation}
For $\chi,\Pi_{\chi}\in I\!\!R$, both $A_{\chi}$ and
$A^{\dag}_{\chi}$ take on all  complex values. Besides,
$\{A_{\chi},A^{\dag}_{\chi}\}=-i$ and
$\bar {A}_{\chi}=A^{\dag}_{\chi}$, the bar denoting complex
conjugation.

The remaining part of the Hamiltonian constraint,
\begin{equation}
h=\frac{1}{2}(\Pi_a^2+a^2+\lambda a^4),
\label{Ha}
\end{equation}
can be regarded as the Hamiltonian of a point particle moving on
the $a$ axis under the influence of the potential $a^2+\lambda
a^4$.  A canonical set of variables in the correponding  phase
space  is given by $h$ and
\begin{eqnarray}
\theta&=&\int^a_{a_{h}}dz(2h-z^2-\lambda z^4)^{-1/2}
\nonumber\\
&=& D^{-1/4}_{h}\;{\rm cn}^{-1}(a_h^{-1}a|\tilde{m}_{h}),
\label{theta}
\end{eqnarray}
where ${\rm cn}^{-1}(u|\tilde{m}_{h})$ is the inverse Jacobian
elliptic function with parameter $\tilde{m}_{h}$ \cite{JAC}, and
$D_{h}$, $a_{h}$ and $m_h=1-\tilde{m}_h$ are the values taken by
the parameters $D$, $a_M$ and $m$ [defined in Eqs.
(3.9,10)] when $E=h$.  It is not difficult to check that
$h$ is the momentum canonically conjugate to $\theta$.

{}From the above equations, it follows that $h\in I\!\!R^+$,
and that $a_{h}$ is the maximum value permitted clasically for
$a$ when the energy of the point particle is $h$.  On the other
hand, taking into account that ${\rm nc}(iu|m)={\rm cn}(u|1-m)$,
Eq. (\ref{theta}) can be seen to provide the analytic
continuation to the Lorentzian regime of the Euclidean classical
solution (\ref{aconf}), with $h$ and $\theta$ substituting for
$E$ and the Lorentzian conformal time, respectively.

Had we neglected the restriction $a\in I\!\!R^+$, Eq.
(\ref{theta}) would have implied that, for $h$ fixed, the scale
factor should describe orbits in phase space which are periodic
in $\theta$, with period
\begin{equation}
4\int_0^{a_{h}}dz(2h-z^2-\lambda z^4)^
{-1/2}=4D_{h}^{-1/4} K(\tilde{m}_{h}),\end{equation}
$K(\tilde{m}_{h})$ denoting again the complete elliptic integral
of the first kind.  However, the restriction to positive scale
factors breaks this periodicity, limiting the classical motion
in the $(a,\Pi_a)$ plane to only half of each periodic orbit.
Since the dynamics is invariant under a flip of sign in $a$, and
we have chosen the origin of $\theta$ at the turning point
$a_{h}$ of the scale factor, we conlude that all allowed
trajectories on phase space can actually be described by letting
$h\in I\!\!R^+$ and
\begin{equation}
\theta\in (-I_{h},I_{h}),
\hspace{5mm}
\mbox{with }\ I_{h}=D^{-1/4}_h K(\tilde{m}_{h}).
\label{thetaIh}
\end{equation}

We can now introduce the annihilation and creation like
variables
\begin{equation}
A_a=\sqrt{h} e^{-i\theta},\;\;\;\;\;\;\, A^{\dag}_a=\sqrt{h}
e^{i\theta}.
\label{As}
\end{equation}
These variables verify $\{A_a,A^{\dag}_a\}=-i$ and
$\bar A_a=A^{\dag}_a$. However, given restriction
(\ref{thetaIh}), their range is not the whole complex plane.
None the less, this will not lead to any problem in the
quantization of the system, because the only physically relevant
conditions on quantum operators reflecting restrictions on the
range of classical variables are those which refer to the
observables of the quantum theory.

The quotient $A^{\dag}_a/A_a=e^{2i\theta}$ distinguishes all
points $\theta\in (-I_h,I_h)$ for  fixed $h$, because $I_h$ can
be shown to be  within the interval $(0,\frac{\pi}{2})$ for
positive $h$.  As a consequence, expressions (\ref{As}) admit
the inversion
\begin{equation}
h=A^{\dag}_a A_a,\hspace{1cm} \theta=-\frac{i}{2}
\;\ln\left(\frac{A^{\dag}_a}{A_a}\right).
\end{equation}
The change of variables from $(\theta,h)$ to $(A_a,A^{\dag}_a)$
is therefore analytic in the whole phase space of the model.

In the following, we shall regard
$(A_{\chi},A^{\dag}_{\chi},A_a,A^{\dag}_a)$ as our complete set
of elementary variables. Notice that this set is indeed closed
both under Poisson brackets and complex conjugation.

Let us define now
\begin{eqnarray}
&N_{\chi}=A^{\dag}_{\chi}A_{\chi},\hspace{13mm}& N_a=A^{\dag}_a
A_a,
\label{Ns}
\\
&J_+=\frac{1}{\sqrt{2}}A^{\dag}_{\chi}A^{\dag}_a,\hspace{8mm}&
J_-=\frac{1}{\sqrt{2}}A_{\chi}A_a.
\label{Js}
\end{eqnarray}
The Hamiltonian constraint (\ref{Hlor}) can then be rewritten as
$H=N_a-N_{\chi}=0$.
Moreover, taking into account that
\begin{equation}
\{A_x,A^{\dag}_x\}=-i,\;\;\;\;\{A_x,N_x\}=-iA_x,\;\;\;\;
\{A^{\dag}_x,N_x\}=iA^{\dag}_x
\label{algebra}
\end{equation}
with $x=\chi,a$, one can check that the variables
(6.9,10) are actually observables of the model,
because their Poisson brackets with $H$ vanish.  Since
$N_{\chi}$ and $N_a$ coincide modulo the constraint $H=0$, we
will restrict all further considerations to the set
$(J_+,J_-,N_{\chi})$. This set of observables can be easily
proved to be (over-)complete.

Given  that $A_{\chi}$ and $A^{\dag}_{\chi}$ can take on any
complex value, the range of $J_+$ and $J_-$ is the whole complex
plane.  Besides, recalling that $\bar {A}_x=A^{\dag}_x$
($x=\chi,a$), we get the reality conditions
\begin{equation}
\bar {J}_+=J_-,\;\;\;\;\;\;
\bar {N}_{\chi}=N_{\chi}\in I\!\!R^{+}.
\label{JJNN}
\end{equation}
Finally, we also have
\begin{equation}
\{J_+,N_{\chi}\}=iJ_+,\hspace{7mm}\{J_-,N_{\chi}\}=-iJ_-,
\end{equation}
\begin{equation}
\{J_+,J_-\}=\frac{i}{2}(N_a+N_{\chi})\approx iN_{\chi},
\label{JJN}
\end{equation}
the last identity holding weakly.  Therefore, the observables
$(J_+,J_-,N_{\chi})$ generate the Lie algebra of $SO(2,1)$ under
Poisson brackets.

\subsection{Representation space}

In order to quantize the system, we should  represent the
elementary classical variables of the model via linear operators
acting on a certain vector space. The space that we shall choose
for this task will be that of complex functions on
$I\!\!R^+\times I\!\!R$ spanned by the basis
\begin{equation}
\psi_{nm}(a,\chi)=\Phi_n(a)\varphi_m(\chi)
\;\;\;\;\;\;(a\in I\!\!R^+,\;\chi\in I\!\!R),
\label{psinm}
\end{equation}
with $n$ and $m$ two arbitrary non-negative integers and
$\varphi_m(\chi)$ the normalized wave functions of the harmonic
oscillator. Here, the functions $\Phi_n(a)$ are the solutions to
Eq. (\ref{wdwa}) which decrease exponentially at  infinity.
We have shown in Sec. V that these functions are analytic in the
semiaxis $a\geq 0$. This and the damped asymptotic behavior
guarantee that the integrals $\int_{I\!\!R^+}da
\bar {\Phi}_n(a) \Phi_n(a)$ converge. We shall assume
hereafter that the functions $\Phi_n(a)$ have been normalized so
that the above integrals are equal to the unity.

Our representation space contains all the wormhole solutions
constructed in Sec. V, namely $\psi_{nn}(a,\chi)$. We finally
want to show that the basis $\psi_{nm}(a,\chi)$  is linearly
independent.  Since the wave functions $\varphi_m(\chi)$ are
known to possess this property, it will suffice to prove the
linear independence of the functions $\Phi_n(a)$, with $a\in
I\!\!R^+$. Let us then suppose that
\begin{equation}
\sum_{s=1}^{p} c_{n_s} \Phi_{n_s}(a)=0,
\label{cphi}
\end{equation}
where $\{n_s\}$ is an ordered set of non-negative integers,
\makebox{$p>1$} is another integer, and the $c_{n_s}$'s are complex
constants.  Acting on both sides of this equation with the
operator
\begin{equation}
\prod_{s=1}^{p-1}\left(\hat{H}_a-n_{s}-\frac{1}{2}\right),
\end{equation}
 in which
$\hat{H}_a$ is  defined in Eq. (\ref{wdwa}), we get
\begin{equation}
c_{n_p}(n_p-n_{p-1})..\cdots (n_p-n_1)\Phi_{n_p}(a)=0.
\end{equation}
We thus  conclude that $c_{n_p}$ must vanish, since
$\Phi_{n_p}(a)\neq 0$ and $n_p>n_s$ for $s=1,\ldots,p-1$.
Substituting now $c_{n_p}=0$ in Eq.  (\ref{cphi}) and iterating
the above procedure, we arrive at $c_n=0$ for all $n\in
\{n_s\}$.  Therefore, the functions $\Phi_n(a)$ on
$I\!\!R^+$ are linearly independent, and so is then the
basis $\psi_{nn}(a,\chi)$ of our representation space.

\subsection{Quantization}

The elementary variables
$(A_{\chi},A^{\dag}_{\chi},A_a,A^{\dag}_a)$ will now be
represented as linear operators on the complex vector space
spanned by the functions $\psi_{nm}(a,\chi)$, where
$n,m=0,1\ldots$ The action of the correponding operators on this basis
will be given by
\begin{equation}
\hat{A}_{\chi}\psi_{nm}=\sqrt{m} \psi_{n(m-1)},
\hspace{.5cm}
\hat{A}^{\dag}_{\chi}\psi_{nm}=\sqrt{m+1} \psi_{n(m+1)},
\label{Apsi3}
\end{equation}
\begin{equation}
\hat{A}_{a}\psi_{nm}=\sqrt{n} \psi_{(n-1)m},
\hspace{.5cm}
\hat{A}^{\dag}_{a}\psi_{nm}=\sqrt{n+1} \psi_{(n+1)m},
\label{Apsi2}
\end{equation}
where we have set again $\hbar=1$.  Let us also introduce the
operators
\begin{equation}
\hat{N}_x=\frac{1}{2}(\hat{A}^{\dag}_x\hat{A}_x+\hat{A}_x
\hat{A}^{\dag}_x)\;\;\;\;\;\;\; (x=\chi,a),
\end{equation}
to represent the derived classical variables (\ref{Ns}). From
the above definitions, we obtain the non-vanishing commutators
\begin{equation} [\hat{A}_x,\hat{A}^{\dag}_x]=\hat{1},\;\;\;[\hat{A}_x,
\hat{N}_x]=\hat{A}_x,\;\;\;[\hat{A}^{\dag}_x,\hat{N}_x]=-\hat{A}^{\dag}_x,
\end{equation}
which reproduce the Poisson bracket algebra (\ref{algebra}) up
to the usual factor $i$. Here, $\hat{1}$ is the identity
operator.

We shall next represent the Hamiltonian constraint by
$\hat{H}=\hat{N}_a-\hat{N}_{\chi}$. Recalling that the functions
$\psi_{nm}(a,\chi)$ are linearly independent, it is then
straightforward to see that all quantum solutions to the
Hamiltonian
constraint  have the form
\begin{equation}
\Psi(a,\chi)=\sum_{n=0}^{\infty}c_n \psi_{nn}(a,\chi),
\label{psiachi}
\end{equation}
where the $c_n$'s are arbitrary complex numbers. The vector
space of quantum states, $V_p$, is thus spanned by the wormhole
wave functions $\psi_{nn}(a,\chi)$.

Defining
\begin{equation}
\hat{J}_+=\frac{1}{\sqrt{2}}\hat{A}^{\dag}_{\chi}
\hat{A}^{\dag}_a,\;\;\;\;\;\;\hat{J}_-=\frac{1}{\sqrt{2}}\hat{A}_{\chi}
\hat{A}_a,
\label{Jaa}
\end{equation}
we get from Eqs. (6.19,20)
\begin{eqnarray}
\hat{J}_+\psi_{nn}&=&\frac{1}{\sqrt{2}}(n+1)
\psi_{(n+1)(n+1)},\\
\hat{J}_-\psi_{nn}&=&\frac{1}{\sqrt{2}}n
\psi_{(n-1)(n-1)},\\
\hat{N}_{\chi}\psi_{nn}&=&(n+\frac{1}{2})
\psi_{nn}=\hat{N}_a\psi_{nn}.
\end{eqnarray}
The above operators are hence quantum observables, for they
leave the space $V_p$ of quantum states invariant.

Notice that $\hat{N}_{\chi}$ and $\hat{N}_a$ coincide on $V_p$
due to the Hamiltonian constraint. On the other hand,
comparison of Eqs. (\ref{Js}) and (\ref{Jaa}) shows that
$\hat{J}_+$ and $\hat{J}_-$ represent the classical observables
$J_+$ and $J_-$. We also have on $V_p$
\begin{equation}
[\hat{J}_+,\hat{J}_-]=-\hat{N}_{\chi},\;\;\;\;
[\hat{J}_+,\hat{N}_{\chi}]=-\hat{J}_+,\;\;\;\;[\hat{J}_-,\hat{N}_{\chi}]=
\hat{J}_-,
\end{equation}
which is the algebra of commutators that follows from the
corresponding Poisson brackets. The vector space $V_p$ carries
then a linear representation of the algebra of physical
observables of the model, namely the Lie algebra of $SO(2,1)$.
This representation is actually irreducible, because all the
elements in the basis $\psi_{nn}(a,\chi)$ of $V_p$ can
be reached from each other through the repeated action of the
observables $\hat{J}_+$ and $\hat{J}_-$.

To determine the inner product on $V_p$, we must impose the
reality conditions (\ref{JJNN}) as adjointness relations between
 quantum observables, i.e.  $\hat{J}_+^{\;\star}=\hat{J}_-$
and $\hat{N}_{\chi}^{\;\star}=\hat{N}_{\chi}$ (the star denoting
the Hermitian adjoint). In addition, since $N_{\chi}\in
I\!\!R^+$, the operator $\hat{N}_{\chi}$ should be positive
on the resulting Hilbert space of physical states.  In fact, the
relation $\hat{J}_+^{\;\star}=\hat{J}_-$ suffices to fix the
following inner product on $V_p$, up to a positive constant
factor:
\begin{equation} \langle\Gamma,\Psi\rangle=\langle\sum_{m=0}^{\infty}
d_m \psi_{mm},
\sum_{n=0}^{\infty} c_n\psi_{nn}\rangle=\sum_{n=0}^{\infty}
\bar {d}_n c_n,
\label{gpsi}
\end{equation}
where we have made use of expression (\ref{psiachi}), valid for
all quantum states.

The completion of the vector space $V_p$ with respect to the
above product  supplies then the physical Hilbert space ${\cal
H}_p$ of the quantum theory.  It is clear from Eq. (\ref{gpsi})
that ${\cal H}_p$ is isomorphic to $l^2$, the space of square
summable sequences. One can also easily check that the
observable $\hat{N}_{\chi}$ is indeed a positive operator on
${\cal H}_p$. So, all the reality conditions on the observables
of the system have been satisfactorily dealt with.

It is worth pointing out that, being $V_p$ spanned by the
wormhole wave functions $\psi_{nn}(a,\chi)$, every physical
state in the Hilbert space ${\cal H}_p$ can be interpreted as a
superposition of quantum wormholes.  The inner product
(\ref{gpsi}) can then be regarded as the one picked out on the space of
wormholes by the
reality conditions.

To close this section, we shall  prove that the product obtained
on $V_p$ can be equivalently written in the form
\begin{equation}
\langle\Gamma,\Psi\rangle=\int_{I\!\!R^+}da\int_{I\!\!R}d\chi
\bar {\Gamma}(a,\chi)\Psi(a,\chi).
\label{gpsi2}
\end{equation}
Given that the eigenstates $\varphi_n(\chi)$ of the harmonic
oscillator form an orthonormal basis of $L^2(I\!\!R,d\chi)$
and that the functions $\Phi_n(a)$ have been chosen to have unit
norm in $L^2(I\!\!R^+,da)$, we get
\begin{eqnarray}
\sum_{m=0}^{\infty}\bar {d}_m\sum_{n=0}^{\infty}c_n
\int_{I\!\!R^+}da\;\bar {\Phi}_m(a)\Phi_n(a)&&\nonumber\\
\times\int_{I\!\!R}d\chi
\;\bar {\varphi}_m(\chi)\varphi_n(\chi)
&=&\sum_{n=0}^{\infty}
\bar {d}_n c_n,
\end{eqnarray}
from what it follows that the right hand sides of Eqs.
(\ref{gpsi}) and (\ref{gpsi2}) actually coincide on $V_p$.

\section{Conclusions}
\setcounter{equation}{0}

Among the topology changes that may take place in asymptotically
large regions, it is of particular interest in cosmology the
study of tunnelling effects mediated by wormholes in
asymptotically anti-de Sitter regions of the universe, in which
the effective cosmological constant is negative. It did not seem
quite clear whether these tunnellings could be consistently
described quantum mechanically or, at least, semiclassically.
In this work, we have shown that it is actually
possible to construct a quantum theory for this kind of topology
changes, at least at the level of a minisuperspace model.

We have considered a homogeneous and isotropic minisuperspace
model with a negative cosmological constant and a conformally
coupled massless scalar field. The classical solutions to the
Euclidean equations of motion and the Hamiltonian constraint are
asymptotically anti-de Sitter wormholes. Such solutions are
parametrized by three arbitrary constants that account for the
initial scale factor and conformal field as well as for the energy
of the conformal field, which must be positive.

Starting with a general analysis in superspace, we have seen
that adding suitable surface terms renders the Euclidean action
finite on classical solutions, while ensuring its gauge
invariance and determining  a well-defined variational problem
consistent with appropriate wormhole boundary conditions.  For
our minisuperspace model, these boundary conditions essentially
amount to identifying the gravitational and conformal field
energies with  an equal fixed value in the asymptotically
anti-de Sitter region.  Since the obtained action is finite on
classical solutions, it could be used to reach a consistent
semiclassical treatment  for the asymptotically anti-de Sitter
wormholes.

Two procedures have been employed in order to quantize our
minisuperspace model. We have first written the path integral in
terms of our  Euclidean action. We have argued that wormhole
wave functions can be obtained from this path integral as the
product of an eigenfunction of the harmonic oscillator for the
conformal field and a wave function for a scale factor
restricted to be positive.

To carry out a thorough and complete quantization of the system we
have then followed Ashtekar's program. Thus, we have represented an
appropriately chosen set of elementary variables as quantum
operators acting on a vector space of functions which contains
the wormhole solutions of the model. The Lorentzian reality
conditions have then enabled us  to determine the physical inner
product. This can be understood as an inner product in the space
of quantum wormholes. All the wormhole wave functions
turn out to have finite norm and, moreover, provide an orthonormal basis of
the space of physical states.

\acknowledgments

We are very grateful to Mariano Moles for helpful
discussions and support. C.B. was supported by a Spanish
Ministry of Education and Science (MEC) grant.  L.J.G. was supported by a
joint fellowship from  MEC  and the British Council.
P.G.-D. acknowledges DGICYT for financial support under Research
Projects Nos. PB94--0107 and PB93--0139.  G.A.M.M. was
supported by funds provided by DGICYT and MEC  under Contract
Adjunct to the Project No. PB93--0139.

\end{document}